# ON THE DESIGN OF EXPERIMENTS FOR THE STUDY OF EXTREME FIELD LIMITS IN THE INTERACTION OF LASER WITH ULTRARELATIVISTIC ELECTRON BEAM

S. V. Bulanov[a], T. Zh. Esirkepov[a], Y. Hayashi[a], M. Kando[a], H. Kiriyama[a], J. K. Koga[a], K. Kondo[a], H. Kotaki[a], A. S. Pirozhkov[a], S. S. Bulanov[b], A. G. Zhidkov[c], P. Chen[d], D. Neely[e], Y. Kato[f], N. B. Narozhny[g], and G. Korn[h,i]

[a]*Kansai Photon Science Institute, JAEA, Kizugawa, Kyoto 619-0215, Japan*
[b]*University of California, Berkeley, CA 94720, USA*
[c]*Osaka University, Osaka 565-0871, Japan*
[d]*Leung Center for Cosmology and Particle Astrophysics of the National Taiwan University, Taipei 10617, Taiwan*
[e]*Central Laser Facility, STFC, Rutherford Appleton Laboratory, Didcot OX11 0QX, United Kingdom*
[f]*The Graduate School for the Creation of New Photonics Industries, Hamamatsu, Shizuoka 431-1202, Japan*
[g]*Moscow Engineering Physics Institute (State University), Moscow 115409, Russia*
[h]*Max-Planck-Institut für Quantenoptik, Garching 85748, Germany*
[i]*ELI Beamline Facility, Institute of Physics, Czech Academy of Sciences, Prague 18221, Czech Republic*

**Abstract.** We propose the experiments on the collision of laser light and high intensity electromagnetic pulses generated by relativistic flying mirrors, with electron bunches produced by a conventional accelerator and with laser wake field accelerated electrons for studying extreme field limits in the nonlinear interaction of electromagnetic waves. The regimes of dominant radiation reaction, which completely changes the electromagnetic wave-matter interaction, will be revealed in the laser plasma experiments. This will result in a new powerful source of ultra short high brightness gamma-ray pulses. A possibility of the demonstration of the electron-positron pair creation in vacuum in a multi-photon processes can be realized. This will allow modelling under terrestrial laboratory conditions neutron star magnetospheres, cosmological gamma ray bursts and the Leptonic Era of the Universe.

# 1. INTRODUCTION

According to a widely accepted point of view, classical and quantum electrodynamics represent well understood and fully complete areas of science whereas one of the forefronts of fundamental physics is in string theory, the predictions of which will hopefully bring experimental physics to novel, higher than ever, levels. In the second half of the 20th century and in the beginning of the 21st century humankind witnessed enormous progress in elementary particle physics with the formulation and experimental proof of the Standard Model and in cosmology with the observational evidence of dark matter and dark energy in the Universe. At the same time classical physics continued its vigorous development which resulted in the understanding of the nature of chaos in simple mechanical systems with its relation to the turbulence problem and in the achievements in nonlinear wave theory, which led to the formation of a novel area in mathematical physics, thus demonstrating that there cannot be a fully complete area of science, as known by prominent physists [1]. In the beginning of the 21st century there appeared a demand to understand the cooperative behaviour of relativistic quantum systems and to probe the quantum vacuum. It was realized that the vacuum probing becomes possible by using high power lasers [2]. With increasing of the laser intensity, we shall encounter novel physical processes such as the radiation reaction dominated regimes and then the regime of the quantum electrodynamics (QED) processes. Near the intensity, corresponding to the QED critical electric field, light can generate electron-positron pairs from vacuum and the vacuum begins to act nonlinearly [3].

There are several ways to achieve the higher intensity required for revealing these processes.

One of the methods is based on the simultaneous laser frequency upshifting and pulse compression. These two phenomena were demonstrated within the Relativistic Flying Mirror (RFM) concept, which uses the laser pulse compression, frequency up-shift, and focusing by counter-propagating breaking plasma waves – relativistic parabolic mirrors [4]. In the proof of principle experiments for this concept a narrow band XUV generation was demonstrated [5] with the high photon number [6].

Another way was demonstrated in large scale experiments [7], where a 50 GeV bunch of electrons from SLAC interacted with a counter-propagating laser pulse of the intensity of approximately of $5\times10^{17}$ W/cm$^2$. Gamma-rays with the energy of 30 GeV produced in the multi-photon Compton scattering then subsequently interacted with the laser light creating the electron-positron pairs. The conclusion on appearance of the gamma-rays with the energy of 30 GeV has been obtained by analising the spectra of scatterred electrons and positrons.

The direct evidence of the gamma-ray emission through the spectral measurements, so far, is of particular importance. The direct measurement similar to [8] would allow the selection of the non-linear processes from the consequent linear photon scattering and, therefore, the quantitative verification of theoretical approaches.

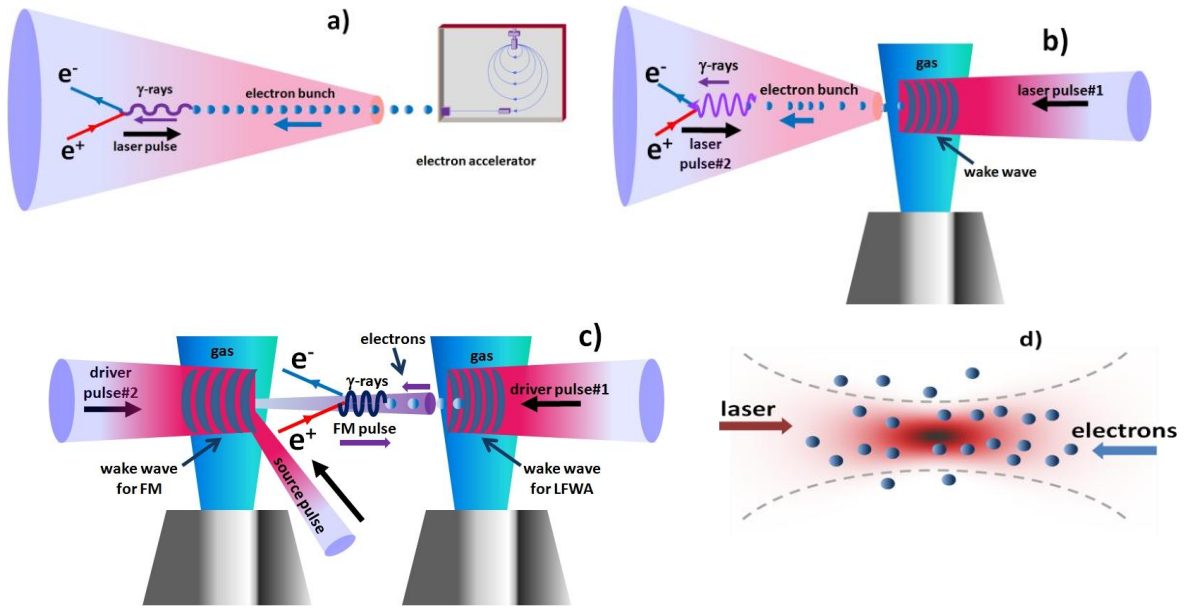

**Figure 1.** Schematic setup of the proposed experiment on the electron-positron pair generation via the Breit-Wheeler process. a) Relativistic electron bunch collides with the tightly focused laser pulse, b) Same as a), but laser pulse collides with the LWFA accelerated electron bunch; in contrast to a) the electron bunch size can be comparable with the laser focus spot size and the Rayleigh length. c) LWFA electron bunch collides with the ultra-high intensity EM wave generated with the relativistic flying mirror. d) Electron bunch interaction with focused laser pulse.

In the present paper we propose table-top experiments on the collision of electromagnetic waves with electron bunch in three configurations, **Fig. 1**. In the first configuration, **Fig. 1 a**, fast electrons are produced by a compact microtron affording hundred MeV bunches with the duration of picosecond. In the second configuration, **Fig. 1 b**, electron bunches are generated in plasma, where a high-intensity laser pulse excites wake waves accelerating electrons [9]. This laser wake field accelerator (LWFA) uses so called self-injected electrons which enter into the accelerating phase of the wake wave due to wave-breaking. LWFA can generate GeV electron bunches with duration of tens femtosecond. In both configurations the electron bunch collides with a petawatt power laser pulse. In particular, ideally the spot size of electron beam should be matched to the focused laser spot, the length matched to the Rayleigh length, and relative timing jitter minimized. Electron bunches generated by the LWFA can be optimally synchronized with the counter-propagating petawatt laser pulse. In the third configuration, **Fig. 1 c**, LFWA-generated bunches collide with an extremely intense electromagnetic pulse produced with the Relativistic Flying Mirror (RFM) technique. The RFM reflecting counter-propagating laser pulse upshifts its frequency and shortens its duration due to the double Doppler effect. It can focus the reflected radiation to the spot much smaller than the laser wavelength due to higher reflected radiation frequency [4]. As we discuss later, in these three configurations electron bunches colliding with counter-propagating laser pulses produce gamma-rays via nonlinear Thomson or inverse multiphoton Compton scattering. The third configuration is designed for the most intense interaction, i.e., for paving a way towards extreme field limits in the nonlinear interaction of electromagnetic waves.

The paper is organized as follows. In the Second section we present the key dimensionless parameters characterizing the extreme field limits. Then the electromagnetic field parameters required for probing the nonlinear vacuum are briefly discussed. The Fourth section contains a consideration of the electron-positron gamma-ray plasma generation via the multi-photon Breit-Wheeler process. Section 5 is devoted to the formulation of possible approaches towards nonlinear vacuum probing and multi-photon electron-positron pair generation with present-day lasers and charged particle accelerator systems. The final Section is devoted to conclusions and discussions of the results obtained.

## 2. EXTREME FIELD LIMITS

Physical systems obey scaling laws, which can also be presented as similarity rules. In the theory of similarity and modeling the key role is played by dimensionless parameters, which characterize the phenomena under consideration [10]. The dimensionless parameters that characterize the high intensity Electromagnetic (EM) wave interaction with matter can also be found in Ref. [11]. The key in the extreme field limit parameters are as follows.

1. Normalized dimensionless EM wave amplitude,

$$a = \frac{eE\lambdabar}{m_e c^2}, \qquad (1)$$

where $\lambdabar = c/\omega = \lambda/2\pi$, corresponds to the intensity $I = 1.37 \times 10^{18} (1\ \mu\text{m}/\lambda)^2 a^2$ W/cm for linear polarization. At $a = 1$, the laser electric field $E$ acting on the electric charge $e$ produces a work equal to $m_e c^2$ over the distance $\lambdabar$. The quiver electron energy becomes relativistic. For plane EM wave this parameter is related to the Lorentz invariant, which being expressed via the 4-potential of the electromagnetic field, $A_\mu$, with $\mu = 0,1,2,3$, is equal to

$$a = \frac{e\sqrt{A_\mu A^\mu}}{m_e c^2}. \qquad (2)$$

It is also the gauge invariant (on the gauge invariance in classical and quantum physics see, e.g. [12]). Below we shall use notations $a_0$ and $a_m$ for initial and maximum values of the normalized electric field, which should not be confused with the zeroth and spacial components of the 4-vector $a_\mu = eA_\mu / m_e c^2$.

2. For the EM emission by an electron the essential parameters are as follows. The first parameter is $\varepsilon_{rad}$, which is proportional to the ratio between the classical electron radius, $r_e = e^2 / m_e c^2 = 2.8 \times 10^{-13}$ cm, and the EM wavelength,

$$\varepsilon_{rad} = \frac{2 r_e}{3 \lambdabar_0}. \qquad (3)$$

Other two parameters are above introduced normalized amplitude, $a$, and the relativistic electron gamma-factor, $\gamma_e$. According to the $\Pi$-theorem of similarity and modeling theory [10], the characteristic dimentionless parameter has a form $\varepsilon_{rad}^x a^y \gamma_e^z$ with the indices $x, y, z$, which finding requires additional consideration. For example, for a circularly polarized EM wave in frame of reference, where the electron is on average on rest, the power emitted by the electron in the ultrarelativistic limit is proportional to the forth power of its energy [13],

$$P_{C,\gamma} = \varepsilon_{rad} \omega_0 m_e c^2 \gamma_e^2 (\gamma_e^2 - 1) \propto \gamma_e^4. \qquad (4)$$

In the non-radiative approximation, the electron can aquire the energy from the EM field with the rate $\approx \omega_0 m_e c^2 a_0$, i.e. $\gamma_e = a$. The condition of the balance between the acquired and lost energy yields $a^3 \approx \varepsilon_{rad}^{-1}$, i.e. the indices $x, y, z$ are $x = 1, y = 3, z = 0$, and the key dimensionless parameter is $a^3 \varepsilon_{rad}$. The radiation effects become dominant at

$$a > a_{rad} = \varepsilon_{rad}^{-1/3}, \qquad (5)$$

i.e. in the limit $I > 2 \times 10^{23}$ W/cm$^2$ for $\lambda = 1 \mu$m [14, 15]. If $\lambda = 1.5 \times 10^{-8}$ cm, which corresponds to a typical wavelength for the XFEL generated EM radiation (see [16]), the radiation effects become significant at $a > a_{rad} \approx 20$, i.e. for $I > 3 \times 10^{28}$ W/cm$^2$. The characteristic frequency of this radiation, $\omega_{\max} = 0.29 \omega_0 \gamma_e^3$, is proportional to the cube of the electron energy [17]; here $\gamma_e = a$. When we consider the electron moving in the plane of null magnetic field formed by two colliding EM waves, $\omega_0$ is the wave frequency.

3. Quantum electrodynamics (QED) effects become important, when the energy of the photon generated by Compton scattering is of the order of the electron energy, i.e. $\hbar \omega_m = \gamma_e m_e c^2$. An electron with energy $\gamma_e m_e c^2$ rotating with frequency $\omega$ in a circularly polarized wave emits photons with energy $\hbar \omega_m \approx \hbar \omega \gamma_e^3$. The quantum effects come into play when $\gamma_e > \gamma_Q = (m_e c^2 / \hbar \omega)^{1/2} \approx 600 (\lambda / 1 \mu m)^{1/2}$, i.e. in terms of the EM field normalized amplitude [15]

$$a_Q = \frac{2e^2 m_e c}{3\hbar^2 \omega} = \frac{2}{3} \alpha \frac{m_e c^2}{\hbar \omega} = \left(\frac{2}{3}\alpha\right)^2 \varepsilon_{rad}^{-1}, \quad (6)$$

where $\alpha = r_e / \lambdabar_C = e^2 / \hbar c \approx 1/137$ is the fine structure constant and $\lambdabar_C = \hbar / m_e c = 3.86 \times 10^{-11}$ cm is the reduced Compton wavelength. Here we take into account that in the limit $I > 10^{23} (1 \mu m / \lambda)^2$ W/cm$^2$ due to strong radiation damping effects the electron energy scales as $m_e c^2 (a / \varepsilon_{rad})^{1/4}$.

4. The limit of the critical QED field, also called the Schwinger field,

$$E_S = \frac{m_e^2 c^3}{e\hbar} = \frac{m_e c^2}{e \lambdabar_C} = 1.32 \times 10^{16} \frac{V}{cm}, \quad (7)$$

corresponding to the intensity of $cE_S^2 / 4\pi \approx 4.6 \times 10^{29}$ W/cm$^2$, is characterized by the normalized EM field amplitude

$$a_S = \frac{m_e c^2}{\hbar \omega} \approx 4.1 \times 10^5 \left(\frac{\lambda}{1\mu m}\right). \quad (8)$$

The electric field $E_S$ acting on the electric charge $e$ produces a work equal to $m_e c^2$ over the distance equal to the Compton wavelength $\lambdabar_C$ [3, 18 - 20].

We note here that the electric field corresponding to the condition $a = a_Q$, where $a_Q$ is given by Eq. (6), is equal to

$$E_Q = \frac{2e^2 m_e c}{3\hbar^2 \omega} = \frac{2}{3}\alpha E_S, \quad (9)$$

and does not depend on the EM wave wavelength with the EM field intensity $I_Q = cE_Q^2 / 4\pi = 3.6 \times 10^{24}$ W/cm$^2$.

When the electric field strength approaches the critical QED limit the electron-positron pair creation from vacuum becomes possible. However, as is well known, the plane EM wave does not create the electron-positron pairs, because this process is determined by the Poincare invariants [3]

$$\mathfrak{F} = \frac{F_{\mu\nu}F^{\mu\nu}}{4} = \frac{\boldsymbol{E}^2 - \boldsymbol{B}^2}{2} \quad \text{and} \quad \mathfrak{G} = \frac{F_{\mu\nu}\varepsilon^{\mu\nu\rho\sigma}F_{\rho\sigma}}{4} = \frac{\boldsymbol{E}\cdot\boldsymbol{B}}{2}, \tag{10}$$

which for plane EM wave vanish. Here $F_{\mu\nu} = \partial_\mu A_\nu - \partial_\nu A_\mu$ is the electromagnetic field tensor expressed via the electromagnetic field 4-vector potential $A_\mu$, and $\varepsilon^{\mu\nu\rho\sigma}$ is the fully antisymmetric unit tensor where $\mu, \nu, \rho, \sigma$ are integers from 0 to 4. For the electron-positron pair creation to occure it requires non plane EM wave configuration.

5. In QED the charged particle interaction with EM fields is determined by the relativistically and gauge invariant parameter

$$\chi_e = \frac{\sqrt{(F^{\mu\nu}p_\nu)^2}}{m_e c E_S}. \tag{11}$$

The parameter $\chi_e$ characterizes the probability of the gamma-photon emission by the electron with Lorentz factor $\gamma_e$ in the field of the EM wave [21 - 24]. It is of the order of the ratio $E/E_S$ in the electron rest frame of reference. Another parameter,

$$\chi_\gamma = \frac{\hbar\sqrt{(F^{\mu\nu}k_\nu)^2}}{m_e c E_S}, \tag{12}$$

is similar to $\chi_e$ with the photon 4-momentum, $\hbar k_\nu$, instead of the electron 4-momentum, $p_\nu$. It characterizes the probability of the electron-positron pair creation due to the collision between the high energy photon and EM field. The parameter $\chi_e$ defined by Eq. (11) can be expressed via the electric and magnetic fields and electron momentum as

$$\chi_e = \frac{1}{E_S m_e c}\sqrt{\left(m_e c \gamma_e \boldsymbol{E} + \boldsymbol{p}\times\boldsymbol{B}\right)^2 - \left(\boldsymbol{p}\cdot\boldsymbol{E}\right)^2}. \tag{13}$$

For the parameter $\chi_\gamma$ defined by Eq. (12) we have

$$\chi_\gamma = \frac{\hbar}{E_S m_e c}\sqrt{\left(\frac{\omega_\gamma}{c}\boldsymbol{E} + \boldsymbol{k}_\gamma\times\boldsymbol{B}\right)^2 - \left(\boldsymbol{k}_\gamma\cdot\boldsymbol{E}\right)^2}. \tag{14}$$

These processes acquire optimal rate at $\chi_e, \chi_\gamma > 1$. For the laser pulse collision with with counterpropagating photon or electron this condition is satisfied at $a > (2/3)\alpha\gamma_{e,\gamma}^{-1}\varepsilon_{rad}^{-1}$.

In order to get into regimes determined by the above-metioned parameters the required laser intensity should be of the order of or above $I > 10^{23}$ W/cm$^2$. This will bring us to experimentally unexplored domain. At such intensities the laser interaction with matter becomes strongly dissipative, due to efficient EM energy transformation into high energy gamma rays [15, 25]. These gamma-photons in the laser field may produce electron-positron pairs via the Breit-Wheeler process [26]. Then the pairs accelerated by the laser generate high energy gamma quanta and so on [22 - 24], and thus the conditions for the avalanche type discharge are produced at the intensity $\approx 10^{24}$ W/cm$^2$. The occurrence of such "showers" was foreseen by Heisenberg and Euler [19].

Relativistic mirrors [4 - 6] may lead to an EM wave intensification resulting in an increase of pulse power up to the level when the electric field of the wave reaches the Schwinger limit when electron-positron pairs are created from the vacuum and the vacuum refractive index becomes nonlinearly dependent on the EM field strength [19, 27]. In quantum field theory particle creation from the vacuum under the action of a strong field has attracted a great amount of attention, because it provides a typical example of non-perturbative processes. In the future, nonlinear QED vacuum properties can be probed with strong and powerful EM pulses.

### 3. PROBING NONLINEAR VACUUM

#### 3.1. Electron-positron pair creation from vacuum

Understanding the mechanisms of vacuum breakdown and polarization is challenging for nonlinear quantum field theories and for astrophysics [28]. Reaching the Schwinger field limit under Earth-like conditions has been attracting a great attention for a number of years. Demonstration of the processes associated with the effects of nonlinear QED will be one of the main challenges for extreme high power laser physics [2].

Vacuum nonlinearity is characterized by the normalized Poincare invariants

$$\mathfrak{f} = \frac{\mathfrak{F}}{E_S^2} = \frac{\boldsymbol{E}^2 - \boldsymbol{B}^2}{2E_S^2} \qquad \text{and} \qquad \mathfrak{g} = \frac{\mathfrak{G}}{E_S^2} = \frac{\boldsymbol{E} \cdot \boldsymbol{B}}{2E_S^2} \ . \qquad (15)$$

Electron-positron pair creation from vacuum by the EM field [18 - 20] is a tunnelling process, which requires the parameter $\gamma_K = 1/a$ to be much less than unity, i.e. $a \gg 1$, and $K_0 = 2m_e c^2 / \hbar \omega \equiv 2a_S$ to be much larger than unity [29], with the number of pairs created per unit volume in unit time given by

$$W_{e^+e^-} = \frac{e^2 E_S^2}{4\pi^2 \hbar^2 c} \mathfrak{e} \mathfrak{b} \coth\left(\frac{\pi \mathfrak{b}}{\mathfrak{e}}\right) \exp\left(-\frac{\pi}{\mathfrak{e}}\right). \tag{16}$$

Here $\mathfrak{e}$ and $\mathfrak{b}$ are the normalized invariant electric and magnetic fields in the frame of reference, where they are parallel:

$$\mathfrak{e} = \sqrt{\sqrt{\mathfrak{f}^2 + \mathfrak{g}^2} + \mathfrak{f}} \quad \text{and} \quad \mathfrak{b} = \sqrt{\sqrt{\mathfrak{f}^2 + \mathfrak{g}^2} - \mathfrak{f}}. \tag{17}$$

When $\mathfrak{e} \to 0$, the pair creation rate $W_{e^+e^-}$ tends to zero. For $\mathfrak{b} \to 0$ we have

$$W_{e^+e^-} = \frac{e^2 E_S^2}{4\pi^3 \hbar^2 c} \mathfrak{e}^2 \exp\left(-\frac{\pi}{\mathfrak{e}}\right). \tag{18}$$

If the EM field varies sufficiently slowly in space and time, i.e. the characteristic scale of this variation is much larger than the characteristic scale of pair production process, which is given by the Compton length, then the field can be considered as constant in each point of 4-volume and the number of pairs is equal to

$$N_{e^+e^-} = \int W_{e^+e^-} d^4 x, \tag{19}$$

where the integration is performed over the 4-volume.

Since the pairs are produced near the maximum of the electric field $\mathfrak{e} = \mathfrak{e}_m$, we express EM field dependence on time and space coordinates locally as

$$\mathfrak{e} \approx \mathfrak{e}_m \left(1 - \frac{x^2}{2\delta x^2} - \frac{y^2}{2\delta y^2} - \frac{z^2}{2\delta z^2} - \frac{t^2}{2\delta t^2}\right), \tag{20}$$

and substitute it into Eq. (18) for the rate $W_{e^+e^-}$; here $\delta x$, $\delta y$, $\delta z$, and $\delta t$ are the characteristic sizes and duration of high-intensity part of EM field. Integrating it over the 4-volume according to Eq. (19) we obtain for the number of pairs [30, 31]

$$N_{e^+e^-} = \frac{c \delta t \delta x \delta y \delta z}{64\pi^4 \lambdabar_C^4} \mathfrak{e}_m^4 \exp\left(-\frac{\pi}{\mathfrak{e}_m}\right). \tag{21}$$

Pair creation requires the first invariant $\mathfrak{F}$ be positive. This condition can be fulfilled in the vicinity of the antinodes of colliding EM waves, or/and in the configuration formed by several focused EM pulses, [31, 32]. Near the focus region this configuration can be modeled by the axially symmetric three-dimentional (3D) electromagnetic field comprised of time-dependent electric and magnetic fields. As is known, in the 3D case the analog of a linearly polarized EM wave is the TM mode, with poloidal electric and toroidal magnetic fields. The EM configurations approximating the EM fields found in Ref. [32] near the field maximum are described by the solution to Maxwell's equations in vacuum and expressed in terms of Bessel functions and associated Legendre polynomials. The EM fields for the TM mode and the first Poincare invariant are shown in **Fig. 2**. The second invariant, $\mathfrak{G} = (\boldsymbol{E} \cdot \boldsymbol{B})$, is equal to zero, i.e. $\mathfrak{b} = 0$ for this EM configuration.

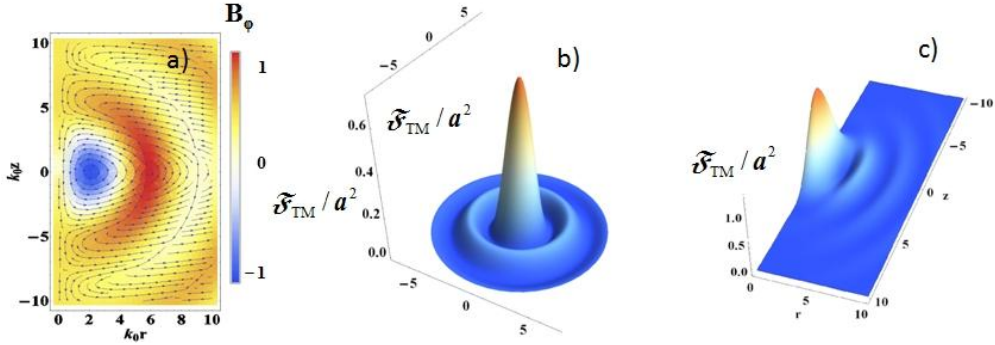

**Figure 2.** a) The vector field shows $r$- and $z$-components of the poloidal electric field in the $r,z$ plane for the TM mode. The color density shows the toroidal magnetic field distribution, $B_\phi(r,z)$. b) The first Poincare invariant $\mathfrak{F}$ normalized by the dimensionless laser amplitude $a_m^2$: $\mathfrak{F}_{TM}(r, z=0, t=0)$. c) $\mathfrak{F}_{TM}(r,z,t=0)$.

For the TM mode, in the vicinity of the electric field maximum, the $z$-component of the electric field oscillates in the vertical direction, the radial component of the electric field is relatively small:

$$\boldsymbol{E} \approx \boldsymbol{e}_z E_m \cos(\omega_0 t) + \boldsymbol{e}_r \frac{E_m}{8} k_0^2 r z \cos(\omega_0 t). \qquad (22)$$

and the $\phi$-component of the magnetic field vanishes on the axis being linearly proportional to the radius,

$$\boldsymbol{B} \approx \boldsymbol{e}_\phi \frac{E_m}{8} k_0 r \sin(\omega_0 t). \qquad (23)$$

Here we use the cylindrical coordinates $r, \phi, z$. Using expression for the probability of electron-positron pair creation given by Eq. (21) and expanding the first Poincare invariant, $\mathfrak{F}_{TM}(r,z,t)$, in the vicinity of its maximum, we find that the pairs are created in a small 4-volume near the electric field maximum with the following characteristic size along the $r$-, $z$-, and $t$-direction:

$$\delta r \approx \lambdabar_0 \left(\frac{5a_m}{\pi a_S}\right)^{1/2}, \quad \delta z \approx \lambdabar_0 \left(\frac{10 a_m}{\pi a_S}\right)^{1/2}, \quad \delta t \approx \omega_0^{-1}\left(\frac{5a_m}{\pi a_S}\right)^{1/2}. \tag{24}$$

Here $a_m = eE_m/m_e\omega_0 c$ is the normalized field amplitude. As shown in [24] the first pairs can be observed for an one-micron wavelength laser intensity of the order of $I^* \approx 10^{27}$ W/cm$^2$, which corresponds to $a_m/a_S = 0.05$, i.e. for a characteristic size approximately equal to $\delta r \approx 0.04 \lambda_0$.

We note here that in the case when the parameter $\gamma_K = 1/a$ characterizing the adiabaticity of the process is large, $\gamma_K = 1/a \gg 1$, i.e. $a \ll 1$, which corresponds to the high frequency EM radiation, and $K = 2 m_e c^2/\hbar\omega$ is large as well, the pair creation occurs in the multiphoton regime with the the number of pairs created per unit volume in unit time given by [29]

$$W_{e^+e^-} \approx \frac{2c}{\pi^3 \lambdabar_C^4 K^{5/2}}\left(\frac{4\gamma_K}{e_N}\right)^{-2K}. \tag{25}$$

Here $e_N = 2.718281828...$ is the Napier- or the Euler number. Assuming the field distribution is given by formula (20) with characteristic size of the order of $\lambdabar$ and integrating Eq. (25) over the 4-volume, we obtain for the number of pairs generated per period in the multiphoton regime

$$N_{e^+e^-} = \frac{\pi}{2^7 K^{9/2}}\left(\frac{e_N a_m}{4}\right)^{2K}. \tag{26}$$

As we see it is always small, provided $a_m \ll 1$, and $K \gg 1$.

### 3.2. Electron-positron gamma-ray plasma generation via the multi-photon Breit-Wheeler process

Reaching the threshold of an avalanche type discharge with electron-positron gamma-ray plasma (EPGP) generation via the multi-photon Breit-Wheeler process [26] discussed in Refs. [21 - 24] requires high enough values of the parameters $\chi_e$ and $\chi_\gamma$ defined above by Eqs. (11) and (12).

In the limit of small parameter $\chi_\gamma \ll 1$ the rate of the pair creation is exponentially small [21],

$$W_\parallel(\chi_\gamma) = \frac{3}{32} \frac{e^2 m_e^2 c^3}{\hbar^3 \omega_\gamma} \left(\frac{\chi_\gamma}{2\pi}\right)^{3/2} \exp\left(-\frac{8}{3\chi_\gamma}\right). \tag{27}$$

In the limit $\chi_\gamma \gg 1$ the pair creation rate is given by

$$W_\parallel(\chi_\gamma) = \frac{27\,\Gamma^7(2/3)}{56\pi^5} \frac{e^2 m_e^2 c^3}{\hbar^3 \omega_\gamma} \left(\frac{3\chi_\gamma}{2}\right)^{2/3}. \tag{28}$$

Here $\hbar\omega_\gamma$ is the energy of the photon with which the Breit-Wheeler process creates an electron-positron pair.

Since for the large $\gamma_e$, the photon is emitted by the electron (positron) in a narrow angle almost parallel to the electron momentum with the energy of the order of the electron energy, the parameters $\chi_e$ and $\chi_\gamma$ are approximately equal to each other, although this is not necessarily so in the limit when the electron emits photons according to classical electrodynamics, i.e. when $\gamma_e < (m_e c^2/\hbar\omega_0)^{1/2} = a_S^{1/2}$. In order to find the threshold for the avalanche development we need to estimate the QED parameter $\chi_e$, defined by Eqs. (11) and (13). The condition for avalanche development corresponding to this parameter is that it should become of the order of unity within one tenth of the EM field period (e.g. see Refs. [22 - 24]). Due to the trajectory bending by the magnetic field the electron transverse momentum changes. It is easy to find the trajectory in the $(r,z)$- plane assuming $\omega_0 t \ll 1$ and $a_m \omega_0 t \gg 1$ of the electron moving in the EM field given by Eqs. (22) and (23). It is described by the relationships

$$p_z(t) = m_e c\, a_m \omega_0 t, \tag{29}$$

$$p_r(t) = m_e c \frac{a_m k_0 r_0 \omega_0 t}{2^{3/2}} I_1\left(\frac{\omega_0 t}{2^{3/2}}\right), \tag{30}$$

$$r(t) = \frac{a_m r_0}{2^{3/2}} I_1\left(\frac{\omega_0 t}{2^{3/2}}\right) + \frac{a_m r_0 \omega_0 t}{16}\left[I_0\left(\frac{\omega_0 t}{2^{3/2}}\right) + I_2\left(\frac{\omega_0 t}{2^{3/2}}\right)\right]. \tag{31}$$

where $I_\nu(x)$ is the modified Bessel function [33] and $r_0$ is the initial electron coordinate, which is of the order of $\delta r$ as given by Eq. (24). It is also assumed that $p_z(0) = p_r(0) = 0$. As we see, the trajectory instability is relatively slow with the growth rate equal to $\omega_0/2^{3/2}$.

According to Eq. (30) the radial component of electron momentum is proportional to the square of time,

$$p_r \approx (a_m/8)k_0\delta r(\omega_0 t)^2. \tag{32}$$

Assuming $\omega_0 t$ to be equal to $0.1\pi$, we obtain from Eq. (12) that $\chi_e$ becomes of the order of unity, i.e. the avalanche can start, at $a_0/a_S \approx 0.08$, which corresponds to the laser intensity $I^* \approx 2.5 \times 10^{27}$ W/cm$^2$. At that limit the Schwinger mechanism provides approximately $10^7$ pairs per one-period of the laser pulse focused to a $\lambda_0^3$ region [31].

In the case of two colliding circularly polarized EM waves, the resulting electric field rotates with frequency $\omega_0$ being constant in magnitude. The radiation friction effects incorporated in the relativistically covariant form of equation of motion of a radiating electron [13] result in

$$m_e c \frac{du^\mu}{ds} = \frac{e}{c} F^{\mu\nu} u_\nu + \frac{2e^2}{3c} g^\mu. \tag{33}$$

Here the radiation friction force in the Lorentz-Abraham-Dirac form is determined by $g^\mu = d^2 u^\mu/ds^2 - u^\mu (du^\nu/ds)(du_\nu/ds)$, $u_\mu = (\gamma, \mathbf{p}/m_e c)$ is the four-velocity, and $s$ is the proper time, $ds = dt/\gamma$.

For the electron rotating in the circularly polarized colliding EM waves the emitted power becomes equal to the maximal energy gain rate at the field amplitude $a > \varepsilon_{rad}^{-1/3}$ (corresponding to the laser intensity $I \approx 4.5 \times 10^{23}$ W/cm$^2$) as was noted above (see Eq. (5)). In this limit the radiation friction effects should be taken into account.

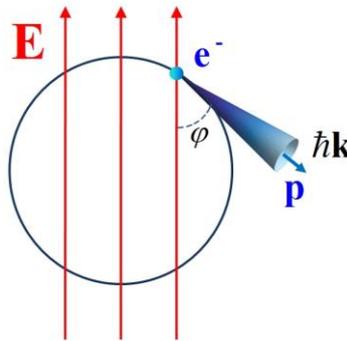

**Figure 3.** Electron moving in the rotating electric field of colliding circularly polarized EM waves emits EM radiation. The angle $\varphi$ between the electron momentum and electric field is determined by Eq. (35).

In order to do this, we represent the electric field and the electron momentum in the complex form:

$$\boldsymbol{E} = E_y + iE_z = E_0 \exp(-i\omega_0 t) \quad \text{and} \quad \boldsymbol{p} = p_y + ip_z = p\exp(-i\omega_0 t - i\varphi), \quad (34)$$

where $\varphi$ is the phase equal to the angle between the electric field vector and the electron momentum (see **Fig. 3**). In the stationary regime the radiation friction force is balanced by the force acting on the electron from the electric field. The electron rotates with constant energy. The equations for the electron energy have the form

$$a_0^2 = (\gamma_e^2 - 1)(1 + \varepsilon_{rad}^2 \gamma_e^6) \quad \text{and} \quad \tan\varphi = \frac{1}{\varepsilon_{rad}\gamma_e^3} \quad (35)$$

with $a_0 = eE_0/m_e\omega_0 c$ and $\gamma_e = \sqrt{1+(p/m_e c)^2}$. In the limit of weak radiation damping, $a_0 \ll \varepsilon_{rad}^{-1/3}$, the absolute value of the electron momentum is proportional to the electric field magnitude,

$$p_\perp = m_e c a_0 \quad \text{with} \quad p_\parallel = m_e c \varepsilon_{rad} a_0^4, \quad (36)$$

i.e. $p_\perp \gg p_\parallel$, while in the regime of dominant radiation damping effects, i.e. at $a_0 \gg \varepsilon_{rad}^{-1/3}$, it is given by

$$p_\parallel = m_e c(a_0/\varepsilon_{rad})^{1/4} \quad \text{and} \quad p_\perp = m_e c(a_0 \varepsilon_{rad})^{-1/2}, \quad (37)$$

i.e. $p_\perp \ll p_\parallel$. Here $p_\perp$ and $p_\parallel$ are the momentum components perpendicular and parallel to the electric field. For the momentum dependence given by this expression the power radiated by the electron is $m_e c^2 \omega_0 a_0$, i.e. the energy obtained from the driving electromagnetic wave is completely re-radiated in the form of high energy gamma rays. At $a_0 \approx \varepsilon_{rad}^{-1/3}$ we have for the gamma photon energy

$$\hbar\omega_\gamma \approx 0.45\hbar \frac{m_e c^3}{e^2}. \quad (38)$$

For example, if $a_0 \approx \varepsilon_{rad}^{-1/3} \approx 450$ the circularly polarized laser pulse with the wavelength of 0.8 μm and intensity of $I \approx 2.2 \times 10^{23}$ W/cm$^2$ generates a burst of gamma photons with the energy of about 20 MeV with the duration determined either by the laser pulse duration or by the decay time of the laser pulse in a plasma.

In the regime when the radiation friction effects are dominant, i.e. when $a_0 \gg \varepsilon_{rad}^{-1/3}$, the angle $\varphi$ between the electron momentum and the electric field is small being equal to $(\varepsilon_{rad} a_0^3)^{-1/4}$, i. e. the electron moves almost opposite to the electric field direction. The electron momentum is given by expression (37), [15, 24]. This yields an estimation

$$\chi_e \approx \frac{a_0 p_\perp}{a_S m_e c} = \left(\frac{a_0}{a_S^2 \varepsilon_{rad}}\right)^{1/2}. \quad (39)$$

This becomes greater than unity for $a_0 > a_S^2 \varepsilon_{rad} \approx 1.6 \times 10^3$, which corresponds to the laser intensity equal to $5.5 \times 10^{24} \text{W/cm}^2$ Without radiation friction taken into account the intensity requirements are much softer. This is the condition of the electron-positron avalanche onset. We see that the radiation friction effects do not prevent the EPGP cascade occurring in the case of circularly polarized colliding laser pulses. Extending our discussion on the XFEL generated photon beams, for which $\varepsilon_{rad} \approx 8 \times 10^{-5}$, we find that the condition $\chi_e > 1$ requires $a_0 > 0.3$. As we see the condition corresponds to the electric field given by Eq. (9) with for which the intensity is equal to $5.5 \times 10^{24}$ W/cm$^2$. The charcteristic field for the avalanche onset is $E_\chi \approx \alpha E_S$, i.e. the quantum radiation effects become dominant at the field strenghth a factor 1/137 smaller than the QED critical field $E_S$.

In order to further clarify the role of radiation friction in the charged particle interaction with a relativistically strong EM field we consider the electron motion withing the framework of approximation corresponding to the Landau-Lifshitz form of the radiation friction force [13, 33] (we notice that the radiation friction in the QED limit is discussed in Ref. [35]),

$$g^\mu = \frac{e}{m_e c^2}\left[\frac{\partial F^{\mu\nu}}{\partial x^\lambda}u_\nu u_\lambda - \frac{e}{m_e c^2}\left(F^{\mu\lambda}F_{\nu\lambda}u^\nu - \left(F_{\nu\lambda}u^\lambda\right)\left(F^{\nu\kappa}u_\kappa\right)u^\mu\right)\right]. \quad (40)$$

Retaining leading in the limit $\gamma_e \to \infty$ term we can write it as

$$\boldsymbol{f} = \varepsilon_{rad}\omega_0 \left(\frac{e}{m_e\omega_0 c}\right)^2 \frac{\boldsymbol{p}}{\gamma_e}\left[\left(\gamma_e \boldsymbol{E} + \frac{\boldsymbol{p}\times\boldsymbol{B}}{m_e c}\right)^2 - \left(\frac{\boldsymbol{p}\cdot\boldsymbol{E}}{m_e c}\right)^2\right]. \quad (41)$$

For the case of two colliding circarly polarized EM waves in the zero magnetic field plane, where the rotating electric field is given by $\boldsymbol{E} = E_y + iE_z = (m_e c\omega_0/e)a\exp(-i\omega_0 t)$, this expression takes the form

$$\boldsymbol{f} = \varepsilon_{rad}\omega_0 \frac{\boldsymbol{p}}{\gamma_e}a^2\left[1+\left(\frac{p_\perp}{m_e c}\right)^2\right]. \quad (42)$$

Although the Landau-Lifshitz approximation for the radiation friction force implies the relative weakness of radiation; asymptotically at $t \to \infty$ for stationary solution it gives the same scaling as given by expression (37), when $a \gg \varepsilon_{rad}^{-1/3}$. In **Fig. 4** we show the results of numerical integration of the electron motion equations. Here a) electron trajectory; b) electron momentum vs time; c) time dependences of perpendicular and parallel to the electric field components of the electron momentum are shown. The upper row corresponds to the dissipationless case with $\varepsilon_{rad} = 0$ and to the normalized electric field equal to $a = 5\times 10^3$. For lower row we have $\varepsilon_{rad} = 1.46\times 10^{-8}$, $a = 5\times 10^3$. As we see, the radiation friction qualitatively changes the electron dynamic. The momentum becomes approximately an order of magnitude less than in the case without dissipation. The perpendicular momentum component, $p_\perp = m_e c(a\varepsilon_{rad})^{-1/2}$, becomes approximately an order of magnitude less than the parallel component.

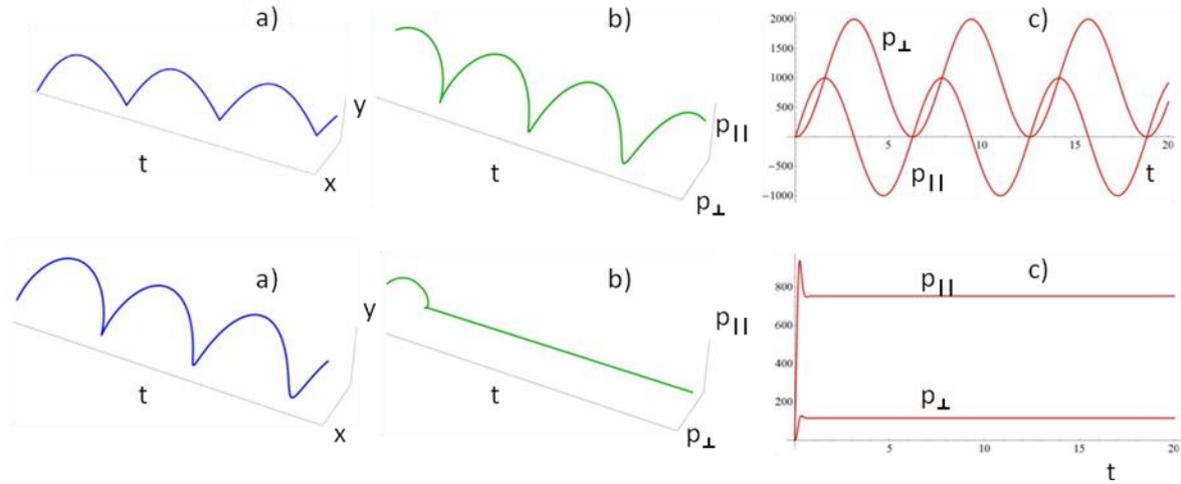

**Figure 4.** Results of numerical integration of the equations of motion for an electron in a circularly polarized electric field: a) electron trajectory; b) electron momentum vs time; c) time dependences of of the electron momentum perpendicular and parallel to the electric field components. Upper row – no radiation friction; $\varepsilon_{rad} = 0$, $a_0 = 5\times 10^3$. Lower row – including the radiation friction; $\varepsilon_{rad} = 1.46\times 10^{-8}$, $a_0 = 5\times 10^3$.

## 4. POSSIBILITY OF NONLINEAR VACUUM PROBING WITH PRESENT - DAY LASER SYSTEMS

### 4.1. Approaching the Schwinger field limit

The concept of the flying mirror has been formulated in Ref. [4] as a way for approaching the critical QED electric field, the Schwinger field limit. This concept is based on the fact that an EM wave reflected off a moving mirror due to the double Doppler effect undergoes frequency multiplication with the multiplication factor $(1+\beta_M)/(1-\beta_M)$ in the limit $\beta_M \to 1$ proportional to the square of the Lorentz factor of the mirror, $\gamma_M = (1-\beta_M^2)^{-1/2}$. This makes this effect an attractive basis for a source of powerful high-frequency radiation. Here $\beta_M = v_M/c$ is calculated for the mirror velocity $v_M$. There are several other schemes for developing compact, intense, brilliant, tunable X-ray sources by using the relativistic mirrors formed in nonlinear interactions in laser plasmas, whose realization will open new ways in nonlinear electrodynamics of continuous media in the relativistic regime (see articles, Refs. [4, 36]).

Here we consider the *relativistic flying mirror* [4 - 6] based on the utilization of the wake plasma waves (**Fig. 1 d**). It uses the fact that the dense shells formed in the electron density in a nonlinear plasma wake, generated by a laser pulse, reflect a portion of a counter-propagating EM pulse. In the wake wave, electron density modulations take the form of a paraballoid moving with the phase velocity close to the speed of light in vacuum [37]. At the wave breaking the electron density in the nonlinear wake wave tends towards infinity. The formation of peaked electron density maxima breaks the geometric optics approximation and provides conditions for the reflection of a substantially high number of photons of the counter propagating EM pulse. As a result of the EM wave reflection from such a "relativistic flying mirror", the reflected pulse is compressed in the longitudinal direction. The paraboloidal form of the mirrors leads to a reflected wave focusing into a spot with the size determined by the shortened wavelength of the reflected radiation.

The key parameter in the problem of the Relativistic Flying Mirror is the wake wave gamma factor, $\gamma_{ph,W}$, which plays a role of the reflecting mirror gamma factor, $\gamma_M$. The number of photons back reflected at the density singularity of the form $n(x) \sim x^{-2/3}$ is proportional to $\approx 0.1 \times \gamma_{ph,W}^{-4}$ (for details see Ref. [38]), which results in the reflected light intensification

$$I/I_0 \approx 13\gamma_{ph,W}^2 \left(S/\lambda_0\right)^2. \tag{43}$$

Here $S$ is the transverse size of the laser pulse efficiently reflected at the RFM. The reflected pulse power is the same as in the incident pulse $\mathcal{P} = \mathcal{P}_0$. If the singularity can be approximated by the Dirac delta-function, $n(x) \sim \delta(x)$, the reflection coefficient is proportional to $\gamma_{ph,W}^{-3}$ with reflected light intensification [3, 30]

$$I/I_0 \approx 32\gamma_{ph,W}^3 (S/\lambda_0)^2. \tag{44}$$

The reflected pulse power increases as $\mathcal{P} = \mathcal{P}_0 \gamma_{ph,W}$.

We note here that utilization of the converging spherical Langmuir wave [39] may be suitable for a flying mirror with higher than (43) and (44) reflected light intensification: $I/I_0 \approx 10^2 \gamma_{ph,W}^5$.

This mechanism allows generating extremely short, femto-, attosecond duration pulses of coherent EM radiation with extremely high intensity.

A demonstration of the Flying Mirror concept has been accomplished in the experiments of Refs. [5]. Two beams of terawatt laser radiation interacted with a gas jet. The first laser pulse excited the nonlinear wake wave in a plasma with the parameters required for the wave breaking. The achievement of this regime was verified by observing the quasi-mono-energetic electron generation and the stimulated Raman scattering. The second counter-crossing laser pulse has been partially reflected from the relativistic mirrors formed by the wake plasma wave. EM pulses with wavelengths from 7 nm to 15 nm and an estimated duration of a few femtoseconds were detected.

The experiments with the 0.5 J, 9 TW laser in the counter-propagating configuration [6] demonstrated dramatic enhancement of the reflected photon number in the extreme ultraviolet wavelength range. The photon number (and reflected pulse energy), is close to the theoretical estimate for the parameters of the experiment.

If the RFM is excited by the pulse of the same amplitude as the incident on it one, then the condition $a_0^2 = 2\gamma_{ph,W}$ follows from the requirement for the wake wave to be in the breaking regime. Using the expression for the reflected pulse intensity (43), we obtain that the electric field in the reflected pulse scales as $E \propto \gamma_{ph,W} E_0$. In order to calculate the threshold for the electron-positron pair creation via the Schwinger mechanism, we take into account that the upshifted frequency is $\omega = 4\gamma_{ph,W}^2 \omega_0$ and reflected pulse is compressed and focused in the region with longitudinal and transverse size equal to $\lambda_0/4\gamma_{ph,W}^2$ and $\lambda_0/2\gamma_{ph,W}$, respectively. Using these relationships and expression (21) we find the number of pairs created by two colliding pulses per one wavelength and one period

$$N_{e^+e^-} = \frac{1}{2^8 \pi^4}\left(\frac{S}{\lambda_0}\right)^4 \exp\left(-\frac{2\pi a_S}{a_0^3}\frac{\lambda_0}{S}\right). \tag{45}$$

From (45) we find that the RFM concept utilization can provide the first pairs to be detected at $a_0 = 100$, i.e. at the source laser intensity of the order of $10^{22}$ W/cm$^2$.

Experiments utilizing the EM pulse intensified with the Relativistic Flying Mirror technique may allow studying regimes of higher-than-Schwinger fields, when $E > E_S$. This may be possible because the light reflected by the parabaloidal FRM is focused into a spot moving with a relativistic velocity and is well collimated within an angle $1/\gamma_{ph,W}$, [4]. The wave localization within the narrow angle means that the wave properties are close to the plane wave properties to the extent of the smallness of the parameter $1/\gamma_{ph,W}$. In this case the first Poincare invariant of the EM field, $\mathfrak{F}$, has a value of the order of $E^2/2\gamma_{ph,W}^2$. The second Poincare invariant, $\mathfrak{G}$ (10), vanishes. Therefore the electric field amplitude in the reflected EM wave can exceed the Schwinger limit by factor $\gamma_{ph,W}$.

We note that a tightly focused EM wave cannot reach an amplitude above $E_S$, due to the electron-positron pair creation [24], which in turn leads to the scattering of the EM wave. While creating and then accelerating the electron-positron pairs the laser pulse generates an electric current and EM field. The electric field induced inside the Lepton-Gamma Plasma (LGP) cloud with a size of the order of the laser wavelength, $\lambda_0$ can be estimated to be $E_{pol} = 2\pi e(n_+ + n_-)\lambda_0$. Here $n_+ \approx n_-$ are the electron and positron density, respectively. When the polarization electric field becomes equal to the laser electric field, $E_{pol} = E_{las}$, the coherent scattering of the laser pulse away from the focus region occurs. This yields for the electron and positron density $n_+ \approx n_- = E_{las}/4\pi e\lambda_0$. The particle number per $\lambda_0^3$ volume is about $a_0\lambda_0/r_e$. This is a factor $a_0$ smaller than that required for the laser energy depletion from the creation of electron-positron pairs, which implies that the laser will be scattered before being depleted.

Due to the nonlinear dependence of the vacuum susceptibilities on the electromagnetic-field amplitude the vacuum index of refraction becomes intensity dependent according to the expression

$$n = 1 + n_2 I, \tag{46}$$

with

$$n_2 I = \frac{\alpha^2}{45\pi}(11 \pm 3)\left(\frac{E}{E_S}\right)^2 \approx 1.6 \times 10^{-37}(11 \pm 3) I \frac{\text{cm}^2}{\text{W}}. \tag{47}$$

Here the EM wave intensity is $I = cE^2/4\pi$. As we see, the nonlinear vacuum susceptibility can approache unity for the laser intensity of the order of $10^{36}$ W/cm$^2$, i.e. higher than the Schwinger limit. The well known Kerr constant of vacuum defined as $K_{QED} = (n-1)/(\lambda_0 |E|^2)$ can be found to be [2, 27]

$$K_{QED} = \left(\frac{7\alpha^2}{90\pi}\right)\left(\frac{\lambdabar_C^3}{m_e c^2 \lambda_0}\right). \tag{48}$$

In order to compare it with the Kerr constant for water, which is equal to $4.7\times10^{-7}$ cm$^2$/erg $(\approx 5\times 10^{-14}$ m$^2$/V$)$, we estimate $K_{QED}$ for the EM wave wavelength $\lambda_0 = 1\,\mu$m and find it be of the order of $8.5\times 10^{-28}$ cm$^2$/erg $(\approx 10^{-34}$ m$^2$/V$)$, i.e. approximately a factor $\approx 10^{20}$ smaller. For the XFEL photon beam, it is smaller by a factor $\approx 10^{12}$.

In the QED nonlinear vacuum two counter-propagating electromagnetic waves mutually focus each other [27]. The critical power $\mathcal{P}_c = cE^2 S_0^2/4\pi$, where $S_0$ is the laser beam waist, for the mutual focusing is equal to

$$\mathcal{P}_{cr} = \frac{45}{14\alpha} cE_S^2 \lambda_0^2, \tag{49}$$

For $\lambda_0 = 1\,\mu$m it yields $\mathcal{P}_c \approx 2.5\times 10^{24}$ W, which is beyond the reach of existing and planned lasers. We notice here that as shorther the EM wavelength, $\lambda_0$, the lower is the critical power. For example, if we take the XFEL wavelength, $\lambda_0 = 1.5\times 10^{-8}$ cm, the critical power corresponds to 50 petawatt limit.

For the laser emitted radiation, if we take into account that the radiation reflected by the RFM has a shortened wavelength, $\lambda_r = \lambda_0/4\gamma_{ph,W}^2$ and that its power is increased by a factor $\gamma_{ph,W}$, we may find that for $\gamma_{ph,W} \approx 30$, i.e. for a plasma density $3\times 10^{17}$ cm$^{-3}$ because $\gamma_{ph,W} = (n_{cr}/n)^{1/2}$, nonlinear vacuum properties can be seen for the laser light incident on the FRM with a power of about 50 PW. This makes the Flying Relativistic Mirror concept attractive for the purpose of studying nonlinear quantum electrodynamics effects. Moreover the collision of two bubbles with or without accelerated electron bunches inside them will provide a good table-top laboratory for studying a number of nonlinear QED effects.

## 4.2. Multi-photon creation of electron-positron gamma-ray plasma in ultrarelativistic electron beam collision with the EM pulse

Here we discuss the requirements for experimental realization of abundant electron-positron pair creation in the PW class laser interaction with relativistic electron beam of moderate energy (see also [40]).

In the experiments [6] on the 527 nm terawatt laser interaction with 46.6 GeV electrons from the SLAC accelerator beam positrons were observed. The positrons were generated in a two-step process in which laser photons were scattered by the electrons and then the generated high-energy photons collided with several laser photons to produce an electron-positron pair. This corresponds to the multiphoton Breit-Wheeler process,

$$N\omega_0 + \omega_\gamma \to e^+ e^-, \qquad (50)$$

where $N$ is the number of laser photons colliding with the gamma-photon to produce the pair.

The parameter $\chi_\gamma$, which determines the probability of the electron-positron pair creation by photons, is given by Eqs. (12) and (14). For a plane EM wave it takes the form

$$\chi_\gamma = \frac{E}{E_S} \frac{\hbar}{m_e c^2} \left( \omega_\gamma - k_{\gamma,x} c \right). \qquad (51)$$

In the counter-propagating configuration we have $\chi_\gamma \approx 2(a/a_S)(\hbar\omega_\gamma/m_e c^2)$.

$$\chi_\gamma \approx 2 \frac{E}{E_s} \frac{\hbar\omega_\gamma}{m_e c^2} = 2a \frac{\hbar^2 \omega_0 \omega_\gamma}{m_e^2 c^4} = 2a \frac{\hbar^2 \bar{\omega}_0 \bar{\omega}_\gamma}{m_e^2 c^4}. \qquad (52)$$

The r.h.s. term of this expression, in which the bars indicate frequencies in another frame of reference, explicitly shows the relativistic invariance of $\chi_\gamma$. It becomes of the order of unity for the energy of the photon, which creates an electron-positron pair, equal to

$$\hbar\omega_\gamma = m_e c^2 \frac{m_e c^2}{2\hbar\omega_0 a}. \qquad (53)$$

Using this expression we can find the number of laser photons $N_l$ required to create the electron-positron pair. In the reference frame where the electron-positron pair is at the rest we have the relationship $N_l \tilde{\omega}_0 = \tilde{\omega}_\gamma$, which gives $N_l = \tilde{\omega}_\gamma / \tilde{\omega}_0 \approx \omega_\gamma / 4\omega_0 \tilde{\gamma}^2$. Here the tilde is used for quantities in the pair rest frame and $\tilde{\gamma}$ corresponds to the Lorentz transformation to this frame. For the photon energy we have $N_l \hbar \tilde{\omega}_0 = m_e c^2$. Since $\hbar \tilde{\omega}_\gamma = m_e c^2$ and $\tilde{\omega}_0 \tilde{\omega}_\gamma = \omega_0 \omega_\gamma = m_e^2 c^4 \chi_\gamma / \hbar^2 a$ we find $\tilde{\gamma} = \chi_\gamma m_e c^2 / 2 \hbar \omega_0 a$, which yields

$$N_l = \frac{a}{\chi_\gamma}. \tag{54}$$

The probability per unit time of the pair creation by photons is given by Eqs. (27) and (28). In the region of $\chi_\gamma \approx 1$, it is of the order of $W \approx 0.1 e^2 m_e^2 c^3 / \hbar^3 \omega_\gamma = 0.1 \alpha \left( m_e c^2 / \hbar \right) \left( \hbar \omega_0 / m_e c^2 \right) a = 0.1 \alpha \omega_0 a$. The mean free path of the photon of the energy $\hbar \omega_\gamma$ before pair creation is equal to $l_{m.f.p} = c / W$,

$$l_{m.f.p} = \frac{\lambda_0}{0.2 \pi \alpha \, a} \approx 220 \frac{\lambda_0}{a}. \tag{55}$$

It is about 2μm for $a = 10^2$.

Now we estimate the parameter $\chi_e$ value, which characterizes the nonlinear Compton scattering. We consider the electron interacting with the plane EM wave. As a result of the interaction with the laser pulse, the electron, being inside the pulse, acquires longitudinal and transverse momentum components. Using the solution presented in Ref. [13], we find that the component of the electron momentum along the direction of the laser pulse propagation can be found from the equation

$$(m_e^2 c^2 + m_e^2 c^2 a^2 + p_x^2)^{1/2} - p_x = (m_e^2 c^2 + p_0^2)^{1/2} - p_0. \tag{56}$$

We use the conservation of generalised momentum which yields for the component of the electron momentum parallel to the laser electric field: $p_y = a m_e c$, provided the radiation friction effects are negligibly small. The $x$-component of the electron momentum before collision with the laser pulse is negative and equal to $-|p_0|$. For a plane EM wave propagating along the x-axis with $\mathbf{E} = E(x - ct) \mathbf{e}_y$ and $\mathbf{B} = E(x - ct) \mathbf{e}_z$, where $E = c^{-1} A'(x - ct)$ and $A(x - ct)$ is the y-component of the 4-vector potential and prime denotes differentiation with respect to the variable $x - ct$, the invariant $\chi_e$ given by Eq. (13) takes the form

$$\chi_e = \frac{E}{E_S} \left( \gamma_e - \frac{p_x}{m_e c} \right). \tag{57}$$

Substituting expression (56) to Eq. (57) we obtain

$$\chi_e = \frac{E}{E_S}\left[\frac{(m_e^2c^2+p_0^2)^{1/2}-p_0}{m_ec}\right] = a\frac{\hbar\omega_0}{m_ec^2}\left[\frac{(m_e^2c^2+p_0^2)^{1/2}-p_0}{m_ec}\right]. \tag{58}$$

For $1\mu$m petawatt laser pulse focused to a few micron focus spot ($a=10^2$) the parameter $\chi_e$ becomes equal to unity for $\gamma_0 = 2.5\times 10^3$, i.e. for the electron energy of about of 1.3 GeV.

The gamma-ray photons depending on the parameters of the laser-electron interaction can be generated either via nonlinear Thomson scattering in the classical electrodynamics regime or via multiphoton inverse Compton scattering, when the quantum mechanical description should be used.

### 4.3. Gamma-ray photon generation via the nonlinear Thomson scattering

At first we analyze nonlinear Thomson scattering. We consider a head-on collision of an ultrarelativistic electron with a laser pulse. The laser ponderomotive pressure pushes the electron sideways with respect to the pulse propagation and also changes the longitudinal component of the electron momentum. In the electron rest frame (see below) the laser pulse duration is $\bar\tau_{las} \approx \tau_{las}\sqrt{1+a^2}/2\gamma_e$. The electron is not scattered aside by the laser ponderomotive force provided its energy is large enough,

$$\gamma_e > c\tau_{las}a/2w_\perp. \tag{59}$$

where $w_\perp$ is the focus width. As we see, for a one-micron, 1 PW, i.e $a=10^2$, 30 fs duration laser pulse focused to an one-wavelength spot this condition requires $\gamma_e > 500$, i.e. the electron energy above 250 MeV. Further we assume that this condition is respected.

When a counter-propagating electron collides with the laser pulse its longitudinal momentum decreases. According to Eq. (56) it is equal to

$$p_x = p_0 + m_ec\frac{a^2 m_ec}{2[(m_e^2c^2+p_0^2)^{1/2}-p_0]}. \tag{60}$$

In the boosted frame of reference where the electron is at the rest the laser frequency is related to the laser frequency in the laboratory frame, $\omega_0$, as [43] (see also Ref. [44] where nonlinear Thomson scattering is discussed)

$$\bar\omega_0 = \frac{\omega_0}{\sqrt{1+a^2}}\sqrt{\frac{1-\beta_0}{1+\beta_0}}, \tag{61}$$

where $\beta_0 = p_0 / \sqrt{m_e^2 c^2 + p_0^2}$. The bar "¯" denotes the frequency value in the frame of reference where the electron on average is at rest. The characteristic frequency of the radiation emitted by the electron in this frame of reference is equal to $\bar{\omega}_m = 0.3 \bar{\omega}_0 a^3$, [17]. The photon energy, $\hbar \bar{\omega}_m = 0.3 \hbar \bar{\omega}_0 a^3$, satisfies condition $\hbar \bar{\omega}_m = m_e c^2 \gamma_e$ for $a = \sqrt{m_e c^2 / 0.3 \hbar \bar{\omega}_0}$, which corresponds to the parameters when the QED description should be used. For lower laser amplitudes the photon generation can be described by the nonlinear Thomson scattering process.

Dividing the radiation power given by Eq. (4) by the typical energy of the emitted photon, $\hbar \bar{\omega}_m = 0.3 \hbar \bar{\omega}_0 a^3$, and multiplying by $2\pi / \bar{\omega}_0$ we find the number of photons emitted by the electron during one wave period,

$$N_\gamma \approx \frac{3\pi}{4} \alpha a. \tag{62}$$

The photon energy in the laboratory frame of reference is

$$\hbar \omega_m = 0.3 \hbar \omega_0 \frac{a^3}{1+a^2} \frac{1-\beta_0}{1+\beta_0} \approx 1.2 \hbar \omega_0 a \gamma_0^2. \tag{63}$$

We note that in the interaction of a GeV electron with a PW laser pulse the radiation friction effects must be incorporated. Here we shall consider it as a perturbation, i.e. in the Landau-Lifshitz form [13], which limits of applicability have been discussed in Refs. [41, 42].

In order to estimate $\omega_\gamma$ we should find the electron beam energy taking into account the radiation losses. Retaining the main order terms in Eq. (40), we obtain the equation for the $x$-component of the electron momentum

$$\frac{dp_x}{dt} = -\varepsilon_{rad} \omega_0 a^2 (2t) \frac{p_x^2}{m_e c}. \tag{64}$$

Its solution is given by

$$p_x(t) = \frac{p_x(0) m_e c}{m_e c + \varepsilon_{rad} \omega_0 p_x(0) \int_0^t a^2(2t') dt'}. \tag{65}$$

If we assume a dependence of $a(t)$ of the form $a(t) = a_0 \exp(-t^2 / 2\tau^2)$, Eq. (65) can be rewritten as

$$p_x(t) = \frac{p_x(0)m_e c}{m_e c + \varepsilon_{rad}\omega_0 p_x(0)\tau a_0^2 \sqrt{\pi/32}\left[\mathrm{erf}\left(\sqrt{2}t/\tau\right)-1\right]}. \qquad (66)$$

Here $\mathrm{erf}(x)$ is the error function equal to $\mathrm{erf}(x) = \left(\sqrt{\pi}/2\right)\int_0^x \exp(-t^2)dt$ [33]. Eq. (66) shows that for large enough $p_x(0)\tau a_0^2$ the electron momentum tends to the limit of

$$p_x(t) \xrightarrow[t\to\infty]{} \sqrt{\pi/8}\left(m_e c / \varepsilon_{rad}\omega_0 \tau a_0^2\right) \qquad (67)$$

in accordance with the theory formulated in Refs. [13,34] and numerical solution of the equations of the electron motion in the laser field with the radiation friction effects taken into account [43]. For $a_0 = 10^2$ and $\omega_0\tau = 6$ the electron momentum is about $p_x(\infty)/m_e c = 500$.

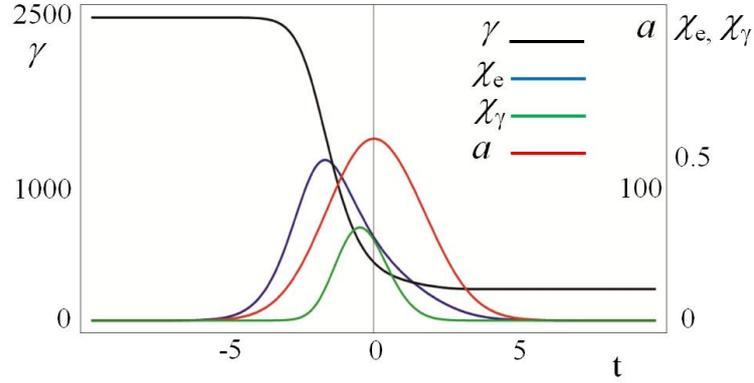

**Figure 5.** Time dependence of the electron and photon parameters for a 1.25 GeV electron beam interacting with a 15 fs PW laser pulse ($a_0 = 100$): the electron energy - black; parameter $\chi_e$ - blue; laser pulse profile – red, parameter $\chi_\gamma$ - green.

In **Fig. 5** we present the time dependence of the electron and photon parameters for a GeV electron beam interaction with the Gaussian PW ($a_0 = 100$, $\tau = 15$ fs) laser pulse, when the ponderomotive force and radiation friction effects are incorporated into the electron equation of motion. The electron energy before and after the interaction with the laser pulse equals 1.25 GeV and 0.25 GeV, respectively. The parameter $\chi_e(t) \approx 2(a(t)/a_S)(p_x(t)/m_e c)$ reaches the maximum value of 0.3 inside the laser pulse. The parameter $\chi_\gamma(t) \approx 2(a(t)/a_S)(\hbar\omega_\gamma(t)/m_e c^2)$ with $\omega_\gamma \approx 0.3\omega_0[(1-\beta_x)/(1+\beta_x)][(p_y^2 + p_z^2)/m_e^2 c^2]^{3/2}$ reaches the maximum value of about 0.3.

## 4.4. Gamma-ray generation via the multi-photon Compton scattering

For multi petawatt lasers with larger laser amplitudes of the focused light, $a > 3m_e c^2 / 2\hbar\omega_0\gamma_0 \approx 300$, the gamma rays are generated in the multiphoton Compton scattering of ultrarelativistic electrons,

$$N\omega_0 + \omega_\gamma \to e^+ e^-. \tag{68}$$

The kinematic consideration of the multiphoton reaction leads to the relationship between the laser frequency, $\omega_0$, and the frequency of scattered photons, $\omega_\gamma$. In the frame of reference, where the electron is at rest, we have (e.g. see [3, 21])

$$\bar{\omega}_\gamma = \frac{N\bar{\omega}_0}{1 + \left(\dfrac{N\hbar\bar{\omega}_0}{m_e c^2} + \dfrac{a^2}{4}\right)(1-\cos\theta)}. \tag{69}$$

where $\theta$ is the angle between the laser pulse and the photon propagation directions, and $a$ is the normalized laser field. At $\theta \to \pi$ the scattered frequency tends to its maximal value $\approx m_e c^2/\hbar$. In the laboratory frame of reference, the scattering cross section is maximal in the narrow angle in the backward direction with respect to the laser propagation.

In the rest frame of the electron the intensity of radiation is given by [21]

$$\frac{dI}{du} = -\frac{e^2 m_e^2 c^3}{4\pi^2 \hbar^2} \frac{u}{(1+u)^3}\left\{\int_z^\infty \text{Ai}(y)dy + \frac{2}{z}\left[1 + \frac{u^2}{2(1+u)}\right]\text{Ai}'(z)\right\}. \tag{70}$$

Here $u = \chi_\gamma / \chi_e$, $z = (u/\chi_e)^{2/3}$, and Ai($z$) is the Airy function. At small $u$ the intensity is proportional to $u^{1/3}$, i.e. $I \sim \bar{\omega}_\gamma^{1/3}$, and at large $u$ it decreases exponentially. The maximum of the intensity distribution is at $\chi_\gamma \approx \chi_e^2$ for $\chi_e \ll 1$ and at $\chi_\gamma \approx \chi_e$ if $\chi_e \gg 1$, i.e. at $\bar{\omega}_\gamma \approx m_e c^2 a/\hbar$.

Using expression (61) for $\bar{\omega}_0$ we obtain that the parameter $\chi_\gamma = a\hbar^2 \bar{\omega}_0 \bar{\omega}_\gamma / m_e^2 c^4$ equals $\chi_\gamma = 2\gamma_0 a\hbar\omega_0 / m_e c^2$. At $a = 10^2$ it becomes of the order of unity for $\gamma_0 \approx 2.5\times 10^3$. We see that the QED consideration gives the same value $\approx 1.25$ GeV for the required relativistic electron energy as in the case of the classical approximation corresponding to nonlinear Thomson scattering.

## 4.5. Electron-positron generation in two laser beams interacting with plasmas

In order to construct a compact source of the electron-positron-gamma-ray plasma it is desirable to produce gamma rays in collisions of the laser accelerated electrons with the EM field of the laser pulse (see **Fig. 1**, where three versions of the experimental setup are presented). As is known, the required GeV range electron energy has been achieved in experiments involving a 40 TW ultrashort laser pulse interaction with underdense plasma [47], when the quasi-mono-energetic electron bunches have been generated in the laser wake field acceleration process.

In the case of usage of the laser wake-field accelerated electrons, the bunch is optimally synchronised with the laser pulse. In addition, the electron density is substantially higher than in the beam of electrons produced by conventional accelerators. For example, a typical microtron accelerator generates a 150 MeV electron beam with the charge of 100 pC and duration of about 20 ps. Being focused to a 50 μm focal spot it provides the electron density $n_b \approx 10^{12} \text{cm}^{-3}$. As a result approximately 300 electrons can interact with the laser pulse. According to Eqs. (54) and (55) the maximal number of electron-positron pairs which can be produced is equal to 300 per shot. However, since for a PW range laser pulse the parameters $\chi_e$ and $\chi_\gamma$ are small compared to unity, the rate of electron-positron pair generation is exponentially low. This requires a number of the laser shots to be accumulated in order to detect the electron-positron pairs.

If we use a LWFA generated electron bunch with a transverse size of several microns, with ten fs duration, e.g. see [45], and with an electric charge of 1 nC, which corresponds to the optimal parameters [46], approximately $10^9$ electrons interact with the laser pulse. For the parameters $\chi_e$ and $\chi_\gamma$ of the order of unity we may expect in the optimal case the generation of approximately $10^9$ pairs per shot. We hote here that in the experiments [47] the electric charge of a 1 GeV electron beam is equal to 30 pC, i.e. the expected number of pairs could be about $3 \times 10^7$ per shot.

Utilization of the EM wave intensified by the relativistic flying mirror to collide with the LWFA accelerated electrons, as is shown in **Fig. 1 c**, can increase further the efficiency of the electron-positron pair generation. In this case the parameters $\chi_e$ and $\chi_\gamma$ increase by a factor equal to $\gamma_{ph,W}$ and can exceed unity, thus providing conditions for the avalanche type prolific generation of the electron-positron pairs.

We note that for the experiments discussed in our paper it is not critical to have high quality electron beams. However, if the discussed gamma-ray sources will find applications in the future, the beam quality will become important depending on specific applications.

Table 1 presents maximum values of the parameters $\chi_e$ and $\chi_\gamma$ for different regimes of the electron bunch interaction with 30 fs PW laser pulses.

Table 1. Peak values of the invariants $\chi_e$ and $\chi_\gamma$ for a 30 fs, PW laser pulse interacting with the RF accelerator generated electron bunch, with the LWFA accelerated electrons, and for the RFM produced EM pulse interacting with the LWFA generated electrons.

|  | RF accelerator | LWFA | LWFA+ RFM EM pulse ($\gamma_{ph,W}=5$) |
|---|---|---|---|
| $\mathcal{E}_e$ | 150 MeV | 1.25 GeV | 1.25 GeV |
| $E/E_S$ | $3\times 10^{-4}$ | $3\times 10^{-4}$ | $1.5\times 10^{-3}$ |
| $\chi_e$ | 0.1 | 0.5 | 2.5 |
| $\chi_\gamma$ | 0.01 | 0.1 | 2.5 |

At present, the optical parametric chirped pulse amplification technique [48] produces $\approx 8\,\text{fs}$, 16 TW laser pulses and several fs petawatt lasers are under development [49]. If we take a 8 fs, 1 PW laser pulse, the parameters $\chi_e$ and $\chi_\gamma$ become equal to 0.16 and 0.018 for a 150 MeV electron beam and 0.7 and 0.25 in the case of the interaction with 1.25 GeV electrons, respectively

**CONCLUSION**

With the concept of the Relativistic Flying Mirror, relatively compact and tunable extremely bright high power sources of ultrashort pulses of x- and gamma-rays become realizable, which will allow for exploring novel physics, for studying the processes of high importance for accelerator physics [50], and for laboratory modeling of processes of key importance for relativistic astrophysics [11, 28]. The experiments in this field will allow modelling under the conditions of a terrestrial laboratory the state of matter in cosmic Gamma Ray Bursts and in the Leptonic Era of the Universe.

## ACKNOWLEDGMENT

We thank G. Dunne, N. Elkina, E. Esarey, D. Habs, T. Heinzl, S. Iso, G. Mourou, M. Murakami, H. Nishimura, H. Ruhl, C. B. Schroeder, T. Tajima, and T. Tauchi for discussions, and appreciate the support from the MEXT of Japan, Grant-in-Aid for Scientific Research (A), 29244065, 2008, and from NSF of USA under Grant No. PHY-0935197.